\documentclass[12pt,onecolumn]{revtex4}
\usepackage{amssymb}
\begin{document}

\title{Charged Rotating Dilaton Black Strings}
\author{M. H. Dehghani$^{1,2}$}\email{mhd@shirazu.ac.ir} \author{N. Farhangkhah $^{1}$}
\address{$1$. Physics Department and Biruni Observatory, Shiraz University, Shiraz 71454, Iran\\
         $2$. Research Institute for Astrophysics and Astronomy of Maragha (RIAAM), Maragha, Iran}

\begin{abstract}
In this paper we, first, present a class of charged rotating solutions in
four-dimensional Einstein-Maxwell-dilaton gravity with zero and
Liouville-type potentials. We find that these solutions can present a black
hole/string with two regular horizons, an extreme black hole or a naked
singularity provided the parameters of the solutions are chosen suitable. We
also compute the conserved and thermodynamic quantities, and show that they
satisfy the first law of thermodynamics. Second, we obtain the ($n+1$%
)-dimensional rotating solutions in Einstein-dilaton gravity with
Liouville-type potential. We find that these solutions can present black
branes, naked singularities or spacetimes with cosmological horizon if one
chooses the parameters of the solutions correctly. Again, we find that the
thermodynamic quantities of these solutions satisfy the first law of
thermodynamics.
\end{abstract}

\maketitle

\section{Introduction\label{Intro}}

The Einstein equation without a cosmological constant has
asymptotically flat
black hole solutions with event horizon being a positive constant curvature $%
n$-sphere. However, for the Einstein equation with positive or negative
cosmological constant \cite{Man, Lem} or Einstein-Gauss-Bonnet equation with
or without cosmological constant \cite{Deh1}, one can have also
asymptotically anti de Sitter (AdS) or de Sitter (dS) black hole solutions
with horizons being zero or negative constant curvature hypersurfaces. These
black hole solutions, whose horizons are not $n$-sphere, are often referred
to as the topological black holes in the literature. Various properties of
these topological black holes have been investigated in recent years. For
example, the thermodynamics of rotating charged solutions of the
Einstein-Maxwell equation with a negative cosmological constant with zero
curvature horizons in various dimensions have been studied in Ref. \cite
{Deh2}, while the thermodynamics of these kind of solutions in Gauss-Bonnet
gravity have been investigated in \cite{Deh3}.

All of the above mentioned black holes are the solutions of field equations
in the presence of a long-range gravitational tensor field $g_{\mu \nu }$
and a long-range electromagnetic vector fields $A_{\mu }$. It is natural
then to suppose the existence of a long-range scalar field too. This leads
us to the scalar-tensor theories of gravity, where, there exist one or
several long-range scalar fields. Scalar-tensor theories are not new, and it
was pioneered by Brans and Dicke \cite{Bran}, who sought to incorporate
Mach's principle into gravity. Also in the context of string theory, the
action of gravity is given by the Einstein action along with a scalar
dilaton field which is non minimally coupled to the gravity \cite{Green}.
The action of ($n+1$)-dimensional dilaton Einstein-Maxwell gravity with one
scalar field $\Phi $ and potential $V(\Phi )$ can be written as \cite{Hor}
\begin{eqnarray}
I_{G} &=&-\frac{1}{16\pi }\int_{\mathcal{M}}d^{n+1}x\sqrt{-g}\left( \mathcal{%
R}\text{ }-\frac{4}{n-1}(\nabla \Phi )^{2}-V(\Phi)-e^{-4\alpha
\Phi/(n-1)}F_{\mu \nu }F^{\mu \nu }\right)  \nonumber \\
&&+\frac{1}{8\pi }\int_{\partial \mathcal{M}}d^{n}x\sqrt{-\gamma }\Theta
(\gamma ),  \label{Act}
\end{eqnarray}
where $\mathcal{R}$ is the Ricci scalar, $\alpha $ is a constant determining
the strength of coupling of the scalar and electromagnetic field, $F_{\mu
\nu }=\partial _{\mu }A_{\nu }-\partial _{\nu }A_{\mu }$ is the
electromagnetic tensor field and $A_{\mu }$ is the vector potential. The
last term in Eq. (\ref{Act}) is the Gibbons-Hawking boundary term. The
manifold $\mathcal{M}$ has metric $g_{\mu \nu }$ and covariant derivative $%
\nabla _{\mu }$. $\Theta $ is the trace of the extrinsic curvature $\Theta
^{\mu \nu }$ of any boundary(ies) $\partial \mathcal{M}$ of the manifold $%
\mathcal{M}$, with induced metric(s) $\gamma _{ij}$. Exact charged dilaton
black hole solutions of the action (\ref{Act}) in the absence of a dilaton
potential [$V(\Phi)=0$] have been constructed by many authors \cite{CDB1,
CDB2}. The dilaton changes the casual structure of the spacetime and leads
to curvature singularities at finite radii. In the presence of
Liouville-type potential [$V(\Phi)=2 \Lambda \exp(2 \beta \Phi)$], static
charged black hole solutions have also been discovered with a positive
constant curvature event horizons and zero or negative constant curvature
horizons \cite{Hor,Cai}. Recently, the properties of these black hole
solutions which are not asymptotically AdS or dS, have been studied \cite
{Clem}.

These exact solutions are all static. Recently, One of us has constructed
two classes of magnetic rotating solutions in four-dimensional
Einstein-Maxwell-dilaton gravity with Liouville-type potential \cite{Deh4}.
These solutions are not black holes, and present spacetimes with conic
singularities. Till now, charged rotating dilaton black hole solutions for
an arbitrary coupling constant has not been constructed. Indeed, exact
rotating black hole solutions have been obtained only for some limited
values of the coupling constant \cite{Fr}. For general dilaton coupling, the
properties of charged dilaton black holes only with infinitesimally small
angular momentum \cite{Hor2} or small charge \cite{Cas} have been
investigated. Our aim here is to construct exact rotating charged dilaton
black holes for an arbitrary value of coupling constant and investigate
their properties.

The outline of our paper is as follows. In Sec \ref{Charged}, we obtain the
four-dimensional charged rotating dilaton black holes/strings which are not
asymptotically flat, AdS or dS, and show that the thermodynamic quantities
of these black strings satisfy the first law of thermodynamics. Section \ref
{Cons} is devoted to a brief review of the general formalism of calculating
the conserved quantities, and investigation of the first law of
thermodynamics for the charged rotating black string. In Sec. \ref{Unch}, we
construct the ($n+1$)-dimensional rotating dilaton black branes, and
investigate their properties. We finish our paper with some concluding
remarks.

\section{Rotating Charged Dilaton Black Strings \label{Charged}}

The field equation of ($n+1$)-dimensional Einstein-Maxwell-dilaton gravity
in the presence of one scalar field $\Phi$ with the potential $V(\Phi)$ can
be written as:
\begin{eqnarray}
&&\partial _{\mu }\left[ \sqrt{-g}e^{-4\alpha \Phi/(n-1)}F^{\mu \nu }%
\right] =0,  \label{Fil1} \\
&&\mathcal{R}_{\mu \nu }=\frac{4}{n-1}\left( \nabla _{\mu }\Phi \nabla _{\nu
}\Phi +\frac{1}{4}Vg_{\mu \nu }\right) +2e^{-4\alpha \Phi/(n-1)%
}\left( F_{\mu \lambda }F_{\nu }^{\text{ }\lambda }-\frac{1}{2(n-1)}F_{\rho
\sigma }F^{\rho \sigma }g_{\mu \nu }\right) ,  \label{Fil2} \\
&&\nabla ^{2}\Phi =\frac{n-1}{8}\frac{\partial V}{\partial \Phi }-\frac{%
\alpha }{2}e^{-4\alpha \Phi/(n-1)}F_{\rho \sigma }F^{\rho \sigma
}. \label{Fil3}
\end{eqnarray}

In this section we want to obtain the four-dimensional charged rotating
black hole solutions of the field equations (\ref{Fil1})-(\ref{Fil3}) with
cylindrical or toroidal horizons. The metric of such a spacetime with
cylindrical symmetry can be written as
\begin{eqnarray}
ds^{2} &=&-f(r)\left( \Xi dt-ad\varphi \right) ^{2}+r^{2}R^{2}(r)\left(
\frac{a}{l^{2}}dt-\Xi d\varphi \right) ^{2}+\frac{dr^{2}}{f(r)}+\frac{r^{2}}{%
l^{2}}R^{2}(r)dz^{2},  \nonumber \\
\Xi ^{2} &=&1+\frac{a^{2}}{l^{2}},  \label{Met1}
\end{eqnarray}
where the constants $a$ and $l$ have dimensions of length and as we will see
later, $a$ is the rotation parameter and $l$ is related to the cosmological
constant $\Lambda $ for the case of Liouville-type potential with constant $%
\Phi $. In metric (\ref{Met1}), the ranges of the time and radial
coordinates are $-\infty <t<\infty $, $0\leq r<\infty $. The topology of the
two dimensional space, $t=$\textrm{constant} and $r=$\textrm{constant}, can
be (\textit{i}) $S^{1}\times S^{1}$ the flat torus $T^{2}$ model, with $%
0\leq \varphi <2\pi $, $0\leq z<2\pi l$, (\textit{ii}) $R\times S^{1}$, the
standard cylindrical symmetric model with $0\leq \varphi <2\pi $, $-\infty
<z<\infty $, and (\textit{iii) }$R^{2}$, the infinite plane model, with $%
-\infty <\varphi <\infty $, $-\infty <z<\infty $ (this planar solution does
not rotate).

The Maxwell equation (\ref{Fil1}) for the metric (\ref{Met1}) is $\partial
_{\mu }\left[ r^{2}R^{2}(r)\exp (-2\alpha \Phi )F^{\mu \nu }\right] =0,$
which shows that if one choose
\begin{equation}
R(r)=\exp (\alpha \Phi ),  \label{Rr}
\end{equation}
then the vector potential can be written as
\begin{equation}
A_{\mu }=-\frac{q}{r}(\Xi \delta _{\mu }^{t}-a\delta _{\mu }^{\varphi }),
\label{met1b}
\end{equation}
where $q$ is the charge parameter. In order to obtain the functions $\Phi
(r) $ and $f(r)$, we write the field equations (\ref{Fil2}) and (\ref{Fil3})
for $a=0$:
\begin{eqnarray}
&&rf^{^{\prime \prime }}+2f^{^{\prime }}(1+\alpha r\Phi ^{^{\prime
}})-2q^{2}r^{-3}e^{-2\alpha \Phi }+rV(\Phi )=0,  \label{E1} \\
&&rf^{^{\prime \prime }}+2f^{^{\prime }}(1+\alpha r\Phi ^{^{\prime
}})+4f[\alpha r\Phi ^{^{\prime \prime }}+2\alpha \Phi ^{\prime }+r(1+\alpha
^{2})\Phi ^{\prime 2}]-2q^{2}r^{-3}e^{-2\alpha \Phi }+rV(\Phi )=0,
\label{E2} \\
&&f^{^{\prime }}(1+\alpha r\Phi ^{^{\prime }})+\alpha f[\rho \Phi ^{^{\prime
\prime }}+2r\alpha \Phi ^{\prime 2}+4\Phi ^{\prime }+(\alpha
r)^{-1}]-q^{2}r^{-3}e^{-2\alpha \Phi }+\frac{1}{2}rV(\Phi )=0,  \label{3} \\
&&f\Phi ^{^{\prime \prime }}+f^{^{\prime }}\Phi ^{^{\prime }}+2\alpha f\Phi
^{\prime 2}+2r^{-1}f\Phi ^{\prime }-\alpha q^{2}r^{-4}e^{-2\alpha \Phi }-%
\frac{1}{4}\frac{\partial V}{\partial \Phi }=0,  \label{E4}
\end{eqnarray}
where the ``prime'' denotes differentiation with respect to $r$. Subtracting
Eq. (\ref{E1}) from Eq. (\ref{E2}) gives:
\[
\alpha r\Phi ^{\prime \prime }+2\alpha \Phi ^{\prime }+r(1+\alpha ^{2})\Phi
^{\prime 2}=0,
\]
which shows that $\Phi (r)$ can be written as:
\begin{equation}
\Phi (r)=\frac{\alpha }{1+\alpha ^{2}}\ln \left( \frac{b}{r}+c\right) ,
\label{Ph}
\end{equation}
where $b$ and $c$ are two arbitrary constants. Using the expression (\ref{Ph}%
) for $\Phi (r)$ in Eqs. (\ref{E1})-(\ref{E4}), we find that these equations
are inconsistent for $c\neq 0$. Thus, we put $c=0$.

\subsection{Solutions with $V(\Phi )=0$}

We begin by looking for the solutions in the absence of a potential ($V(\Phi
)=0$). In this case, it is easy to solve the field equations (\ref{E1})-(\ref
{E4}). One obtains
\[
f(r)=r^{\gamma -1}\left( -C+\frac{(1+\alpha ^{2})q^{2}}{V_{0}r}\right) ,
\]
where $C$ is an arbitrary constant, $\gamma =2\alpha ^{2}/(1+\alpha ^{2})$
and $V_{0}=b^{\gamma }$. In the absence of a non-trivial dilaton ($\alpha
=0=\gamma $), this solution does not exhibit a well-defined spacetime at
infinity. Indeed, $\alpha $ should be greater than or equal to one. In order
to study the general structure of these solutions, we first look for the
curvature singularities. It is easy to show that the Kretschmann scalar $%
R_{\mu \nu \lambda \kappa }R^{\mu \nu \lambda \kappa }$ diverges
at $r=0$, it is finite for $r\neq 0$ and goes to zero as
$r\rightarrow \infty $. Thus, there is an essential singularity
located at $r=0$. Also, it is notable to mention that the Ricci
scaler is finite every where except at $r=0$, and goes to zero as
$r\rightarrow \infty $. The spacetime is asymptotically flat for
$\alpha =1$, while it is neither asymptotically flat nor (A)dS for
$\alpha
>1 $. This spacetime presents a naked
singularity with a regular cosmological horizon at
\[
r_{c}=\frac{(1+\alpha ^{2})q^{2}}{CV_{0}},
\]
provided $C>0$, and no cosmological horizon for $C<0$.

\subsection{Solutions with Liouville-type potentials}

Now we consider the solutions of Eqs. (\ref{E1})-(\ref{E4}) with a
Liouville-type potential $V(\Phi )=2\Lambda \exp (2\beta \Phi )$. One may
refer to $\Lambda $ as the cosmological constant, since in the absence of
the dilaton field $\Phi $ the action (\ref{Act}) reduces to the action of
Einstein-Maxwell gravity with cosmological constant. The only case that we
find exact solutions for an arbitrary values of $\Lambda $ with $R(r)$ and $%
\Phi (r)$ given in Eqs. (\ref{Rr}) and (\ref{Ph}) is when $\beta =\alpha $.
It is easy then to obtain the function $f(r)$ as
\begin{equation}
f(r)=r^{\gamma }\left( \frac{\Lambda V_{0}(1+\alpha ^{2})^{2}}{(\alpha
^{2}-3)}r^{2(1-\gamma )}-\frac{m}{r}+\frac{(1+\alpha ^{2})q^{2}}{V_{0}r^{2}}%
\right) .  \label{Fr1}
\end{equation}
In the absence of a non-trivial dilaton ($\alpha =0=\gamma $), the solution
reduces to the asymptotically AdS and dS charged rotating black string for $%
\Lambda =-3/l^{2}$ and $\Lambda =3/l^{2}$ respectively \cite{Deh2, Lem}.
As one can see from Eq. (\ref{Fr1}), there is no solution for $%
\alpha =\sqrt{3}$ with a Liouville potential ($\Lambda \neq 0$). In order to
investigate the casual structure of the spacetime, we consider it for
different ranges of $\alpha $ separately.

For $\alpha >\sqrt{3}$, as $r$ goes to infinity the dominant term is the
second term, and therefore the spacetime has a cosmological horizon for
positive values of the mass parameter, despite the sign of the cosmological
constant $\Lambda $.

For $\alpha <\sqrt{3}$ and large values of $r$, the dominant term is the
first term, and therefore there exist a cosmological horizon for $\Lambda >0$%
, while there is no cosmological horizons if $\Lambda <0$ . Indeed, in the
latter case ($\alpha <\sqrt{3}$ and $\Lambda <0$) the spacetimes associated
with the solution (\ref{Fr1}) exhibit a variety of possible casual
structures depending on the values of the metric parameters $\alpha $, $m$, $%
q$, and $\Lambda $. One can obtain the casual structure by finding the roots
of $f(r)=0$. Unfortunately, because of the nature of the exponents in (\ref
{Fr1}), it is not possible to find explicitly the location of horizons for
an arbitrary value of $\alpha $. But, we can obtain some information by
considering the temperature of the horizons.

One can obtain the temperature and angular velocity of the horizon by
analytic continuation of the metric. The analytical continuation of the
Lorentzian metric by $t\rightarrow i\tau $ and $a\rightarrow ia$ yields the
Euclidean section, whose regularity at $r=r_{h}$ requires that we should
identify $\tau \sim \tau +\beta _{h}$ and $\varphi \sim \varphi +i\beta
_{h}\Omega _{h}$, where $\beta _{h}$ and $\Omega _{h}$ are the inverse
Hawking temperature and the angular velocity of the horizon. It is a matter
of calculation to show that
\begin{eqnarray}
T_{h} &=&\frac{f^{\text{ }^{\prime }}(r_{h})}{4\pi \Xi }=\frac{r_{h}^{\gamma
-3}}{4\pi \Xi V_{0}(1+\alpha ^{2})}\left[ (3-\alpha
^{2})V_{0}mr_{h}-4(1+\alpha ^{2})q^{2}\right]  \nonumber \\
&=&-\frac{(1+\alpha ^{2})r_{h}^{\gamma -3}}{4\pi \Xi V_{0}}\left[ \Lambda
V_{0}^{2}r_{h}^{4-2\gamma }+q^{2}\right] ,  \label{Tem} \\
\Omega _{h} &=&\frac{a}{\Xi l^{2}}.  \label{Om1}
\end{eqnarray}
Equation (\ref{Tem}) shows that the temperature is negative for the two
cases of (\emph{i}) $\alpha >\sqrt{3}$ despite the sign of $\Lambda $, and (%
\emph{ii}) positive $\Lambda $ despite the value of $\alpha $. As we argued
above in these two cases we encounter with cosmological horizons, and
therefore the cosmological horizons have negative temperature. Numerical
calculations shows that the temperature of the event horizon goes to zero as
the black hole approaches the extreme case. Thus, one can see from Eq. (\ref
{Tem}) that there exist extreme black holes only for negative $\Lambda $ and
$\alpha <\sqrt{3}$, if $r_{h}=4(1+\alpha ^{2})q^{2}/[m_{crit}(3-\alpha
^{2})V_{0}]$, where $m_{crit}$ is the mass of extreme black hole. If one
substitutes this $r_{h}$ into the equation $f(r)=0$, then one obtains the
condition for extreme black string as:
\begin{equation}
m_{crit}=\frac{4(1+\alpha ^{2})q^{2}}{V_{0}(3-\alpha ^{2})}\left( -\frac{%
\Lambda V_{0}^{2}}{q^{2}}\right) ^{(1+\alpha ^{2})/4}.  \label{Mc}
\end{equation}
Indeed, the metric of Eqs. (\ref{Met1}) and (\ref{Fr1}) has two inner and
outer horizons located at $r_{-}$ and $r_{+}$, provided the mass parameter $%
m $ is greater than $m_{crit}$. We will have an extreme black string in the
case of $m=m_{crit}$, and a naked singularity if $m<m_{crit}$. Note that Eq.
(\ref{Mc}) reduces to the critical mass obtained in Ref. \cite{Deh2} in the
absence of dilaton field.

Since the area law of entropy is universal, and applies to all kinds of
black holes and black strings in Einstein gravity, the entropy per unit
length of the string is
\begin{equation}
\mathcal{S}=\frac{\pi \Xi V_{0}r_{h}^{2(1+\alpha ^{2})^{-1}}}{2l},
\label{Ent1}
\end{equation}
where $r_{h}$ is the horizon radius. One may note that the entropy of the
extreme black hole is $\pi \Xi q/(2l\sqrt{-\Lambda })$, which is independent
of the coupling constant $\alpha $.

Finally, it is worthwhile to mention about the asymptotic behavior
of these spacetimes. The asymptotic form of the metric given by
Eqs. (\ref{Met1}) and
(\ref{Fr1}) for the nonrotating case with no cosmological horizon ($\alpha <%
\sqrt{3}$ and $\Lambda <0$) is:
\[
ds^{2}=-\frac{\Lambda V_{0}(1+\alpha ^{2})^{2}}{^{2}(\alpha ^{2}-3)}%
r^{2/(1+\alpha ^{2})}dt^{2}+\frac{(\alpha ^{2}-3)}{\Lambda V_{0}(1+\alpha
^{2})^{2}}r^{-2/(1+\alpha ^{2})}dr^{2}+r^{2/(1+\alpha ^{2})}\left( d\varphi
^{2}+\frac{dz^{2}}{l^{2}}\right) .
\]
Note that $g_{_{tt}}$, for example, goes to infinity as
$r\rightarrow \infty $, but with a rate slower than that of AdS
spacetimes. Indeed, the form of the Ricci and Kretschmann
($\mathcal{K}$) scalars for large values of $r$ are:
\begin{eqnarray*}
\mathcal{R} &=&\frac{6(\alpha ^{2}-2)V_{0}}{(\alpha ^{2}-3)}\Lambda
r^{-2\alpha ^{2}/(1+\alpha ^{2})}, \\
\mathcal{K} &=&\frac{12(\alpha ^{4}-2\alpha ^{2}+2)V_{0}^{2}}{(\alpha
^{2}-3)^{2}}\Lambda ^{2}r^{-4\alpha ^{2}/(1+\alpha ^{2})}.
\end{eqnarray*}
As one may note, these quantities go to zero as $r\rightarrow
\infty $, but with a slower rate than those of asymptotically flat
spacetimes and do not approach nonzero constants as in the case of
asymptotically AdS spacetimes.

\section{The Conserved Quantities and First law of Thermodynamics \label%
{Cons}}

The conserved charges of the string can be calculated through the use of the
substraction method of Brown and York \cite{BY}. Such a procedure causes the
resulting physical quantities to depend on the choice of reference
background. For asymptotically (A)dS solutions, the way that one deals with
these divergences is through the use of counterterm method inspired by
(A)dS/CFT correspondence \cite{Mal}. However, in the presence of a
non-trivial dilaton field, the spacetime may not behave as either dS ($%
\Lambda >0$) or AdS ($\Lambda <0$). In fact, it has been shown that with the
exception of a pure cosmological constant potential, where $\beta =0$, no
AdS or dS static spherically symmetric solution exist for Liouville-type
potential \cite{Pol}. But, as in the case of asymptotically AdS spacetimes,
according to the domain-wall/QFT (quantum field theory) correspondence \cite
{Sken}, there may be a suitable counterterm for the stress energy tensor
which removes the divergences. In this paper, we deal with the spacetimes
with zero curvature boundary [$R_{abcd}(\gamma )=0$], and therefore the
counterterm for the stress energy tensor should be proportional to $\gamma
^{ab}$. Thus, the finite stress-energy tensor in $(n+1)$-dimensional
Einstein-dilaton gravity with Liouville-type potential may be written as
\begin{equation}
T^{ab}=\frac{1}{8\pi }\left[ \Theta ^{ab}-\Theta \gamma ^{ab}+\frac{n-1}{l_{%
\mathrm{eff}}}\gamma ^{ab}\right] ,  \label{Stres}
\end{equation}
where $l_{\mathrm{eff}}$ is given by
\begin{equation}
l_{\mathrm{eff}}^{2}=\frac{(n-1)^{3}\beta ^{2}-4n(n-1)}{8\Lambda }e^{-2\beta
\Phi }.  \label{leff}
\end{equation}
As $\beta $ goes to zero, the effective $l_{\mathrm{eff}}$ of Eq. (\ref{leff}%
) reduces to $l=n(n-1)/2\Lambda $ of the (A)dS spacetimes. The first two
terms in Eq. (\ref{Stres}) is the variation of the action (\ref{Act}) with
respect to $\gamma ^{ab}$, and the last term is the counterterm which
removes the divergences. One may note that the counterterm has the same form
as in the case of asymptotically AdS solutions with zero curvature boundary,
where $l$ is replaced by $l_{\mathrm{eff}}$. To compute the conserved
charges of the spacetime, one should choose a spacelike surface $\mathcal{B}$
in $\partial \mathcal{M}$ with metric $\sigma _{ij}$, and write the boundary
metric in ADM form:
\[
\gamma _{ab}dx^{a}dx^{a}=-N^{2}dt^{2}+\sigma _{ij}\left( d\varphi
^{i}+V^{i}dt\right) \left( d\varphi ^{j}+V^{j}dt\right) ,
\]
where the coordinates $\varphi ^{i}$ are the angular variables
parameterizing the hypersurface of constant $r$ around the origin, and $N$
and $V^{i}$ are the lapse and shift functions respectively. When there is a
Killing vector field $\mathcal{\xi }$ on the boundary, then the quasilocal
conserved quantities associated with the stress tensors of Eq. (\ref{Stres})
can be written as
\begin{equation}
\mathcal{Q}(\mathcal{\xi )}=\int_{\mathcal{B}}d^{n-1}\varphi \sqrt{\sigma }%
T_{ab}n^{a}\mathcal{\xi }^{b},  \label{charge}
\end{equation}
where $\sigma $ is the determinant of the metric $\sigma _{ij}$, $\mathcal{%
\xi }$ and $n^{a}$ are the Killing vector field and the unit normal vector
on the boundary $\mathcal{B}$ . For boundaries with timelike ($\xi =\partial
/\partial t$) and rotational ($\varsigma =\partial /\partial \varphi $)
Killing vector fields, one obtains the quasilocal mass and angular momentum
\begin{eqnarray}
M &=&\int_{\mathcal{B}}d^{n-1}\varphi \sqrt{\sigma }T_{ab}n^{a}\xi ^{b},
\label{Mastot} \\
J &=&\int_{\mathcal{B}}d^{n-1}\varphi \sqrt{\sigma }T_{ab}n^{a}\varsigma
^{b},  \label{Angtot}
\end{eqnarray}
provided the surface $\mathcal{B}$ contains the orbits of $\varsigma $.
These quantities are, respectively, the conserved mass, angular and linear
momenta of the system enclosed by the boundary $\mathcal{B}$. Note that they
will both be dependent on the location of the boundary $\mathcal{B}$ in the
spacetime, although each is independent of the particular choice of
foliation $\mathcal{B}$ within the surface $\partial \mathcal{M}$.

The mass and angular momentum per unit length of the string when the
boundary $\mathcal{B}$ goes to infinity can be calculated through the use of
Eqs. (\ref{Mastot}) and (\ref{Angtot}),
\begin{equation}
\mathcal{M}=\frac{V_{0}}{8l}\left( \frac{(3-\alpha ^{2})\Xi ^{2}+\alpha
^{2}-1}{(1+\alpha ^{2})}\right) m,\hspace{0.5cm}\mathcal{J}=\frac{(3-\alpha
^{2})V_{0}}{8l(1+\alpha ^{2})}\Xi ma.  \label{MJ}
\end{equation}
For $a=0$ ($\Xi =1$), the angular momentum per unit length vanishes, and
therefore $a$ is the rotational parameter of the spacetime. Note that Eq. (%
\ref{MJ}) is valid only for $\alpha <\sqrt{3}$, which the spacetime has no
cosmological horizons. Of course, one may note that these conserved charges
reduce to the conserved charges of the rotating black string obtained in
Ref. in \cite{Deh2} as $\alpha \rightarrow 0$.

Next, we calculate the electric charge of the solutions. To determine the
electric field we should consider the projections of the electromagnetic
field tensor on special hypersurfaces. The normal to such hypersurfaces for
the spacetimes with a longitudinal magnetic field is
\[
u^{0}=\frac{1}{N},\text{ \ }u^{r}=0,\text{ \ }u^{i}=-\frac{V^{i}}{N},
\]
and the electric field is $E^{\mu }=g^{\mu \rho }\exp (-2\alpha \Phi
)F_{\rho \nu }u^{\nu }$. Then the electric charge per unit length $\mathcal{Q%
}$ can be found by calculating the flux of the electric field at infinity,
yielding
\begin{equation}
\mathcal{Q}=\frac{\Xi q}{2l}.  \label{chden}
\end{equation}
The electric potential $U$, measured at infinity with respect to the
horizon, is defined by \cite{Cal}
\[
U =A_{\mu }\chi ^{\mu }\left| _{r\rightarrow \infty }-A_{\mu }\chi ^{\mu
}\right| _{r=r_{+}},
\]
where $\chi =\partial _{t}+\Omega \partial _{\varphi }$ is the null
generators of the event horizon. One obtains
\begin{equation}
U =\frac{q}{\Xi r_{+}}.  \label{Pot}
\end{equation}

Finally, we consider the first law of thermodynamics for the black string.
Although it is difficult to obtain the mass $\mathcal{M}$ as a function of
the extensive quantities $\mathcal{S}$, $\mathcal{J}$ and $\mathcal{Q}$ for
an arbitrary values of $\alpha $, but one can show numerically that the
intensive thermodynamic quantities, $T$, $\Omega $ and $U$ calculated above
satisfy the first law of thermodynamics,

\begin{equation}
dM=Td\mathcal{S}+\Omega d\mathcal{J}+Ud\mathcal{Q}.  \label{Flth}
\end{equation}

\section{The Rotating Black Branes in Various Dimensions \label{Unch}}

In this section we look for the rotating uncharged solutions of field
equations (\ref{Fil1})-(\ref{Fil3}) in $(n+1)$ dimensions with
Liouville-type potential. we first obtain the static solution and then
generalize it to the case of rotating solution with all the rotation
parameters.

\subsection{Static Solutions}

The metric of a static ($n+1$)-dimensional spacetime with an ($n-1$%
)-dimensional flat submanifold $dX^{2}$ can be written as
\begin{equation}
ds^{2}=-f(r)dt^{2}+\frac{dr^{2}}{f(r)}+\frac{r^{2}}{l^{2}}e^{2\beta \Phi
}dX^{2}.  \label{Met2}
\end{equation}
The field equations (\ref{Fil1})-(\ref{Fil3}) for the above metric become
\begin{eqnarray}
&&\frac{r}{n-1}f^{^{\prime \prime }}+f^{^{\prime }}(1+\beta r\Phi ^{^{\prime
}})+\frac{4\Lambda r}{(n-1)^{2}}e^{2\beta \Phi }=0,  \label{EE1} \\
&&\frac{r}{n-1}f^{^{\prime \prime }}+f^{^{\prime }}(1+\beta r\Phi ^{^{\prime
}})+\frac{4\Lambda r}{(n-1)^{2}}e^{2\beta \Phi }  \nonumber \\
&&\hspace{1.6cm}+2(n-2)rf\left[ \beta r\Phi ^{^{\prime \prime }}+2\beta \Phi
^{^{\prime }}+r\left( \beta ^{2}+\frac{4}{(n-1)^{2}}\right) \Phi ^{^{\prime
2}}\right] =0,  \label{EE2} \\
&&f^{^{\prime }}(1+\beta r\Phi ^{^{\prime }})+\beta f[r\Phi ^{^{\prime
\prime }}+(n-1)(2\Phi ^{^{\prime }}+\beta r\Phi ^{^{\prime }2})+(n-2)(\beta
r)^{-1}]+\frac{2\Lambda r}{n-1}e^{2\beta \Phi }=0,  \label{EE3} \\
&&f\Phi ^{^{\prime \prime }}+f^{^{\prime }}\Phi ^{^{\prime }}+(n-1)\left[
\beta f\Phi ^{^{\prime 2}}+r^{-1}f\Phi ^{^{\prime }}-\frac{1}{2}\beta
\Lambda e^{2\beta \Phi }\right] =0.  \label{EE4}
\end{eqnarray}
Subtracting Eq. (\ref{EE1}) from Eq. (\ref{EE2}) gives
\[
\beta r\Phi ^{^{\prime \prime }}+2\beta \Phi ^{^{\prime }}+r\left( \beta
^{2}+\frac{4}{(n-1)^{2}}\right) \Phi ^{^{\prime 2}}=0,
\]
which indicates that $\Phi (r)$ can be written as:
\begin{equation}
\Phi (r)=\frac{(n-1)^{2}\beta }{4+(n-1)^{2}\beta ^{2}}\ln \left( \frac{c}{r}%
+d\right) ,  \label{Ph1}
\end{equation}
where $c$ and $d$ are two arbitrary constants. Substituting $\Phi (r)$ of
Eq. (\ref{Ph1}) into the field equations (\ref{EE1})-(\ref{EE4}), one finds
that they are consistent only for $d=0$. Putting $d=0$, then $f(r)$ can be
written as
\begin{eqnarray}
&&f(r)=\frac{8\Lambda V_{0}}{(n-1)^{3}\beta ^{2}-4n(n-1)}r^{2\Gamma
}-mr^{1-(n-1)\Gamma },  \nonumber \\
&&V_{0}=\Gamma ^{-2}c^{2(1-\Gamma )},\hspace{0.5cm}\Gamma =4\{(n-1)^{2}\beta
^{2}+4\}^{-1}.  \label{Fr2}
\end{eqnarray}
One may note that there is no solution for $(n-1)^{2}\beta
^{2}-4n=0$. It is worthwhile to note that the Ricci scalar goes to
zero as $r\rightarrow \infty $.

The Kretschmann scalar diverges at $r=0$, it is finite for $r\neq
0$, and goes to zero as $r\rightarrow \infty $. Thus, there is an
essential singularity located at $r=0$. Again, the spacetime is
neither asymptotically flat nor (A)dS, but has a regular event
horizon for negative $\Lambda $ at:
\begin{equation}
r_{h}=\left\{ \frac{[(n-1)^{2}\beta ^{2}-4n]m}{\Lambda V_{0}}\right\}
^{1/[(n-1)\Gamma -1]},  \label{rh}
\end{equation}
provided $(n-1)^{2}\beta ^{2}-4n<0$. If $(n-1)^{2}\beta ^{2}-4n>0$, then the
spacetime has a cosmological horizon with radius given in Eq. (\ref{rh}) for
positive values of $\Lambda $. The Hawking temperature of the event or
cosmological horizon is:
\[
T_{h}=-\frac{\Lambda \Gamma V_{0}}{2\pi \Xi }r_{h}^{2\Gamma -1},
\]
which is positive for event horizon ($\Lambda<0$) and negative for
cosmological horizon ($\Lambda>0$).

\subsection{Rotating solutions with all the rotation parameters}

The rotation group in $(n+1)$-dimensions is $SO(n)$ and therefore the number
of independent rotation parameters for a localized object is equal to the
number of Casimir operators, which is $[n/2]\equiv k$, where $[n/2]$ is the
integer part of $n/2$. The generalization of the metric (\ref{Met2}) with
all rotation parameters is
\begin{eqnarray}
ds^{2} &=&-f(r)\left( \Xi dt-{{\sum_{i=1}^{k}}}a_{i}d\phi _{i}\right) ^{2}+%
\frac{r^{2}}{l^{4}}e^{2\beta \Phi }{{\sum_{i=1}^{k}}}\left( a_{i}dt-\Xi
l^{2}d\phi _{i}\right) ^{2}  \nonumber \\
&&-\frac{r^{2}}{l^{2}}e^{2\beta \Phi }{\sum_{i=1}^{k}}(a_{i}d\phi
_{j}-a_{j}d\phi _{i})^{2}+\frac{dr^{2}}{f(r)}+\frac{r^{2}}{l^{2}}e^{2\beta
\Phi }dX^{2},  \nonumber \\
\Xi ^{2} &=&1+\sum_{i=1}^{k}\frac{a_{i}^{2}}{l^{2}},  \label{Met3}
\end{eqnarray}
where $a_{i}$'s are $k$ rotation parameters, $f(r)$ is given in Eq. (\ref
{Fr2}), and $dX^{2}$ is now the Euclidean metric on the $(n-k-1)$%
-dimensional submanifold with volume $V_{n-k-1}$.

The conserved mass and angular momentum per unit volume $V_{n-k-1}$ of the
solution calculated on the boundary $\mathcal{B}$ at infinity can be
calculated through the use of Eqs. (\ref{Mastot}) and (\ref{Angtot}),
\begin{eqnarray}
\mathcal{M} &=&(2\pi )^{k}\frac{\Gamma ^{n}V_{0}^{(n-1)/2}}{16\pi l^{n-k-1}}%
\left[ n-1+\left( n-\frac{(n-1)^{2}\beta ^{2}}{4}\right) (\Xi ^{2}-1)\right]
m, \\
\hspace{0.5cm}\mathcal{J}_{i} &=&(2\pi )^{k}\frac{\Gamma ^{n}V_{0}^{(n-1)/2}%
}{16\pi l^{n-k-1}}\left( n-\frac{(n-1)^{2}\beta ^{2}}{4}\right) \Xi ma_{i}.
\end{eqnarray}

The entropy per unit volume $V_{n-k-1}$ of the black brane is
\begin{equation}
\mathcal{S}=(2\pi )^{k}\frac{\Xi (\Gamma
^{2}V_{0})^{(n-1)/2}r_{h}^{(n-1)\Gamma }}{4l^{n-k-1}},  \label{Ent2}
\end{equation}
where $r_{h}$ is the horizon radius. Again, it is a matter of calculation to
show that the thermodynamic quantities calculated in this section satisfy
the first law of thermodynamics,
\begin{equation}
d\mathcal{M}=Td\mathcal{S}+{{{\sum_{i=1}^{k}}}}\Omega _{i}d\mathcal{J}_{i},
\end{equation}
where $\Omega _{i}=(\Xi l^{2})^{-1}a_{i}$ is the $i$th component of angular
velocity of the horizon.

\section{Closing Remarks}

Till now, no explicit rotating charged black hole solutions have been found
except for some dilaton coupling such as $\alpha =\sqrt{3}$ \cite{Fr} or $%
\alpha =1$ when the string three-form $H_{abc}$ is included \cite{Sen}. For
general dilaton coupling, the properties of charged dilaton black holes have
been investigated only for rotating solutions with infinitesimally small
angular momentum \cite{Hor2} or small charge \cite{Cas} . In this paper we
obtained a class of charged rotating black hole solutions with zero and
Liouville-type potentials. We found that these solutions are neither
asymptotically flat nor (A)dS. In the case of $V(\Phi )=0$, the solution
(which includes the string theoretical case, $\alpha =1$) presents a black
string with a regular event horizon, provided the charge parameter does not
vanish. This solution has not inner horizons, and is acceptable only for $%
\alpha \geq 1$. Thus, it has not a counterpart for Einstein gravity without
dilaton ($\alpha =0$).

In the presence of Liouville-type potential, we obtained exact solutions
provided $\beta =\alpha \neq \sqrt{3}$. These solutions reduce to the
charged rotating black string of \cite{Lem, Deh2}. We found that these
solutions have a cosmological horizon for (\emph{i}) $\alpha >\sqrt{3}$
despite the sign of $\Lambda $, and (\emph{ii}) positive values of $\Lambda $%
, despite the magnitude of $\alpha $. For $\alpha <\sqrt{3}$, the solutions
present black strings with outer and inner horizons if $m>m_{crit}$, an
extreme black hole if $m=m_{crit}$, and a naked singularity if $m<m_{crit}$.
The Hawking temperature of all the above horizons were computed. We found
that the Hawking temperature is negative for inner and cosmological
horizons, and it is positive for outer horizons. We also computed the
conserved and thermodynamics quantities of the four-dimensional rotating
charged black string, and found that they satisfy the first law of
thermodynamics.

Next, we constructed the rotating uncharged solutions of ($n+1$)-dimensional
dilaton gravity with all rotation parameters. If $(n-1)^{2}\beta ^{2}-4n<0$,
then these solutions present a black branes for $\Lambda <0$, and a naked
singularity for $\Lambda >0$. If $(n-1)^{2}\beta ^{2}-4n>0$, the solutions
have a cosmological horizon for positive $\Lambda $, while they are not
acceptable for negative values of $\Lambda $. Again we found that the
thermodynamic quantities of the black brane solutions satisfy the first law
of thermodynamics.

Note that the $(n+1)$-dimensional rotating solutions obtained here are
uncharged. Thus, it would be interesting if one can construct rotating
solutions in $(n+1)$ dimensions in the presence of electromagnetic field.
One may also attempt to generalize these kinds of solutions obtained here to
the case of two-term Liouville potential.
\acknowledgments{This work has been supported by
Research Institute for Astronomy and Astrophysics of Maragha,
Iran}

\end{document}